\documentclass[aps,prl,twocolumn,floatfix,nofootinbib,showpacs,superscriptaddress]{revtex4-1}

\usepackage{amsmath,amsfonts,amssymb,bm}
\usepackage{dsfont}
\usepackage{graphicx}
\usepackage{color}
\usepackage{pzccal}
\usepackage{soul}
\definecolor{purple}{rgb}{0.5,0,0.5}
\definecolor{blue}{rgb}{0.0,0,0.9}

\begin{document}

\title{Pion electromagnetic form factor at spacelike momenta}
\author{L.~Chang}
\affiliation{Institut f\"ur Kernphysik, Forschungszentrum J\"ulich, D-52425 J\"ulich, Germany}

\author{I.\,C.~Clo\"et}
\affiliation{Physics Division, Argonne National Laboratory, Argonne, Illinois 60439, USA}

\author{C.\,D.~Roberts}
\affiliation{Physics Division, Argonne National Laboratory, Argonne, Illinois 60439, USA}

\author{S.\,M.~Schmidt}
\affiliation{Institute for Advanced Simulation, Forschungszentrum J\"ulich and JARA, D-52425 J\"ulich, Germany}

\author{P.\,C.~Tandy} \affiliation{Center for Nuclear Research, Department of
Physics, Kent State University, Kent, Ohio 44242, USA}

\date{12 September 2013}

\begin{abstract}
%
A novel method is employed to compute the pion electromagnetic form factor, $F_\pi(Q^2)$, on the entire domain of spacelike momentum transfer using the Dyson-Schwinger equation (DSE) framework in quantum chromodynamics (QCD).  The DSE architecture unifies this prediction with that of the pion's valence-quark parton distribution amplitude (PDA).  Using this PDA, the leading-order, leading-twist perturbative QCD result for $Q^2 F_\pi(Q^2)$ underestimates the full computation by just 15\% on $Q^2\gtrsim 8\,$GeV$^2$, in stark contrast with the result obtained using the asymptotic PDA.  The analysis shows that hard contributions to the pion form factor dominate for $Q^2\gtrsim 8\,$GeV$^2$ but, even so, the magnitude of $Q^2 F_\pi(Q^2)$ reflects the scale of dynamical chiral symmetry breaking, a pivotal emergent phenomenon in the Standard Model.
\end{abstract}

\pacs{
13.40.Gp    
14.40.Be 	
12.38.Lg    
12.38.Aw    
}

\maketitle

%
\noindent\emph{1:Introduction}.\,---\,The pion occupies a special place in nuclear and particle physics.  It is the archetype for meson-exchange forces \cite{Yukawa:1935xg} and hence, even today, plays a critical role as an elementary field in the nuclear structure Hamiltonian \cite{Pieper:2001mp,Epelbaum:2008ga,Machleidt:2011zz}.  On the other hand, following introduction of the constituent-quark model \cite{GellMann:1964nj,Zweig:1981pd}, the pion came to be considered as an ordinary quantum mechanical bound-state of a constituent-quark and constituent-antiquark.  In that approach, however, explaining its properties requires a finely tuned potential \cite{Godfrey:1985xj}.


The modern paradigm views the pion in a very different manner \cite{Maris:1997hd}: it is both a conventional bound-state in quantum field theory and the Goldstone mode associated with dynamical chiral symmetry breaking (DCSB) in QCD, the strong interaction sector of the Standard Model.
Given this apparent dichotomy, fine tuning should not play any role in a veracious explanation of pion properties.  The pion's peculiarly low (lepton-like) mass, its strong couplings to baryons, and numerous other characteristics are all unavoidable consequences of chiral symmetry and the pattern by which it is broken in the Standard Model.  Therefore, descriptions of the pion within frameworks that cannot faithfully express symmetries and their breaking patterns (such as constituent-quark models) are unreliable.

The fascination of the pion is compounded by the existence of exact results for both soft and hard processes.  For example, there are predictions for low-energy $\pi \pi$ scattering \cite{Weinberg:1966kf,Colangelo:2001df} and the neutral-pion's two-photon decay \cite{Adler:1969gk,Bell:1969ts}; and, on the other hand, perturbative QCD (pQCD) yields predictions for pion elastic and transition form factors at asymptotically high energies \cite{Farrar:1979aw,Efremov:1979qk,Lepage:1980fj}.  The empirical verification of the low-energy results \cite{Batley:2010zza,Larin:2010kq} is complemented by a determined experimental effort to test the high-energy form-factor predictions \cite{Volmer:2000ek,Horn:2006tm,Huber:2008id,Aubert:2009mc,Uehara:2012ag,E1206101}.  In contrast to the low-energy experiments, however, which check global symmetries and breaking patterns that might be characteristic of a broad class of theories, the high-energy experiments are a direct probe of QCD itself; and some would argue that QCD has not passed these tests.

We do not share this view, given that QCD's failure was also suggested in connection with measurements of the pion's valence-quark distribution function \cite{Conway:1989fs} and that those claims are now known to be erroneous \cite{Hecht:2000xa,Wijesooriya:2005ir,Holt:2010vj,Aicher:2010cb,Nguyen:2011jy}.  Nevertheless, an explanation is required for the mismatch between extant experiments on the pion's electromagnetic form factor and what is commonly presumed to be the prediction of pQCD.

The QCD prediction can be stated succinctly \cite{Farrar:1979aw,Efremov:1979qk,Lepage:1980fj}:
\begin{equation}
\label{pionUV}
\exists Q_0>\Lambda_{\rm QCD} \; |\;  Q^2 F_\pi(Q^2) \stackrel{Q^2 > Q_0^2}{\approx} 16 \pi \alpha_s(Q^2)  f_\pi^2 \mathpzc{w}_\varphi^2,
\end{equation}
where $f_\pi=92.2\,$MeV is the pion decay constant \cite{Beringer:1900zz},
\begin{equation}
\alpha_s(Q^2) = 4 \pi/[\beta_0\,\ln(Q^2/\Lambda^2_{\rm QCD})],
\end{equation}
$\beta_0 = 11 - (2/3) n_f$ ($n_f$ is the number of active quark flavours), is the leading-order expression for the strong running coupling, and
\begin{equation}
\label{wphi}
\mathpzc{w}_\varphi = \frac{1}{3} \int_0^1 dx\, \frac{1}{x} \varphi_\pi(x)\,,
\end{equation}
where $\varphi_\pi(x)$ is the pion's valence-quark parton distribution amplitude (PDA).
The value of $Q_0$ is not predicted by pQCD.  (Here $\Lambda_{\rm QCD} \sim 0.2\,$GeV is the natural mass-scale of QCD, whose dynamical generation through quantisation spoils the conformal invariance of the classical massless theory \cite{Collins:1976yq,Nielsen:1977sy,tarrach}.)

Notably, $\mathpzc{w}_\varphi=1$ if one uses the ``asymptotic'' PDA \cite{Farrar:1979aw,Efremov:1979qk,Lepage:1980fj}
\begin{equation}
\label{phiasy}
\varphi_\pi(x) = \varphi^{\rm asy}_\pi(x) = 6 x(1-x).
\end{equation}
This form of the PDA is certainly valid on the domain $\Lambda_{\rm QCD}^2/Q^2\simeq 0$.  As explained elsewhere \cite{Cloet:2013tta}, however, the domain $\Lambda_{\rm QCD}^2/Q^2\simeq 0$ corresponds to \emph{very} large values of $Q^2$.  
This is highlighted by the fact that $\varphi^{\rm asy}_\pi(x)$ can only be a good approximation to the pion's PDA when it is accurate to write $u_{\rm v}^\pi(x) \approx \delta(x)$, where $u_{\rm v}^\pi(x)$ is the pion's valence-quark distribution function.
This is far from valid at momentum scales now accessible \cite{Hecht:2000xa,Wijesooriya:2005ir,Holt:2010vj,Aicher:2010cb,Nguyen:2011jy}.

The perceived disagreement between experiment and QCD theory is based on an observation that at $Q^2=4\,$GeV$^2$, approximately the midpoint of the domain accessible at next-generation facilities \cite{Dudek:2012vr}, Eqs.\,\eqref{pionUV}--\eqref{phiasy} yield
\begin{equation}
\label{pionUV4}
Q^2 F_\pi(Q^2) \stackrel{Q^2=4\,{\rm GeV}^2}{=} 0.15\,,
\end{equation}
where we have used $n_f=4$ and $\Lambda_{\rm QCD}=0.234\,$GeV for illustration \cite{Qin:2011dd}.  The result in Eq.\,\eqref{pionUV4} is a factor of $2.7$ smaller than the empirical value quoted at $Q^2 =2.45\,$GeV$^2$ \cite{Horn:2006tm,Huber:2008id}: $0.41^{+0.04}_{-0.03}$; and a factor of three smaller than that computed at $Q^2 =4\,$GeV$^2$ in Ref.\,\cite{Maris:2000sk}.  Notably, Ref.\,\cite{Maris:2000sk} provided the only prediction for the pointwise behaviour of $F_\pi(Q^2)$ that is both applicable on the entire spacelike domain currently mapped reliably by experiment and confirmed thereby.

In this case the perception of a mismatch and a real discrepancy are not equivalent because, as indicated above, one can convincingly argue that $Q^2=4\,$GeV$^2$ is not within the domain $\Lambda_{\rm QCD}^2/Q^2\simeq 0$ upon which Eq.\,\eqref{phiasy} is valid \cite{Cloet:2013tta}.  This being so and given the successful prediction in Ref.\,\cite{Maris:2000sk}, one is naturally led to ask whether the methods used therein can address the issue of the ultimate validity of Eq.\,\eqref{pionUV}.

Until recently, the answer was ``no'', owing to an over-reliance hitherto on brute numerical methods in such computations.  That has now changed, however, with a refinement of known methods \cite{Nakanishi:1963zz,Nakanishi:1969ph,Nakanishi:1971} described recently in association with a computation of the pion's light-front wave-function \cite{Chang:2013pq}.  As we illustrate herein, these methods enable reliable computation of the pion's electromagnetic form factor to arbitrarily large-$Q^2$ and the correlation of that result with Eq.\,\eqref{pionUV} using the consistently computed distribution amplitude, $\varphi_\pi(x)$.

%
\noindent\emph{2:Computing the pion form factor}.\,---\,At leading order in the systematic and symmetry-preserving DSE truncation scheme introduced in Refs.\,\cite{Munczek:1994zz,Bender:1996bb} and reviewed in Refs.\,\cite{Chang:2011vu,Bashir:2012fs}, the pion form factor is given by
\begin{eqnarray}
\nonumber
K_\mu F_\pi(Q^2) & = & N_c {\rm tr}_{\rm D} 
\int\! \frac{d^4 k}{(2\pi)^4}\,
\chi_\mu(k+p_f,k+p_i) \\
&& \times \Gamma_\pi(k_i;p_i)\,S(k)\,\Gamma_\pi(k_f;-p_f)\,, \quad\label{RLFpi}
\end{eqnarray}
where $Q$ is the incoming photon momentum, $p_{f,i} = K\pm Q/2$, $k_{f,i}=k+p_{f,i}/2$, and the remaining trace is over spinor indices. 
The other elements in Eq.\,\eqref{RLFpi} are the dressed-quark propagator
\begin{equation}
S(p) = -i \gamma\cdot p \, \sigma_V(p^2,\zeta^2)+\sigma_S(p^2,\zeta^2)\,,
\end{equation}
which, consistent with Eq.\,\eqref{RLFpi}, is computed from the rainbow-truncation gap equation ($\zeta$ is the renormalisation scale); the pion Bethe-Salpeter amplitude $\Gamma_\pi(k;P)$,
%
computed in rainbow-ladder truncation; and the unamputated dressed-quark-photon vertex, $\chi_\mu(k_f,k_i)$, which should also be computed in rainbow-ladder truncation. 
[The impact of corrections to the leading-order (rainbow-ladder) computation is understood.  The dominant effect is a modification of the power associated with the logarithmic running in Eq.\,\eqref{pionUV}.  That running is slow and hence the diagrams omitted have no material impact on the discussion herein.]

The leading-order DSE result for the pion form factor is now determined once an interaction kernel is specified for the rainbow gap equation.  In common with Ref.\,\cite{Chang:2013pq}, we use the kernel explained in Ref.\,\cite{Qin:2011dd}.  The strength of this interaction is specified by a product: $D\omega = m_G^3$.  With $m_G$ fixed, results for properties of ground-state vector and flavour-nonsinglet pseudoscalar mesons are independent of the value of $\omega \in [0.4,0.6]\,$GeV \cite{Qin:2011xq}.  We use $\omega =0.5\,$GeV.  With this kernel, $f_\pi=0.092\,$GeV is obtained with $m_G(\zeta=2\,{\rm GeV})=0.87\,$GeV.

By using precisely the rainbow-ladder kernel described in Ref.\,\cite{Chang:2013pq}, we are spared the need to solve numerically for the dressed-quark propagator and pion Bethe-Salpeter amplitude.  Instead, we can employ the generalised Nakanishi representations for $S(p)$ and $\Gamma_\pi(k;P)$ described therein.

That is not the case for $\chi_\mu(k_f,k_i)$, however, because such a representation is not yet available.  We therefore use the following \emph{Ansatz}, expressed solely in terms of the functions which characterise the dressed-quark propagator ($q=k_f-k_i$)
\begin{eqnarray}
\nonumber
\chi_\mu(k_f,k_i) & = & \gamma_\mu X_1(k_f,k_i) + \gamma\cdot k_{f}\gamma_{\mu}\gamma\cdot k_{i}  \, X_2(k_f,k_i) \\
\nonumber
&& +  \, i \, [\gamma\cdot k_f \gamma_\mu + \gamma_\mu \gamma\cdot k_i ] \, X_3(k_f,k_i) \\
&& - \, \tilde\eta\, \sigma_{\mu\nu} q_\nu \, \sigma_S(q^2)\, X_1(k_f,k_i)\,, \label{ChiAnsatz}
\end{eqnarray}
where
$\tilde\eta$ is a parameter and, with $\Delta_{F}(k_f^2,k_i^2)= [F(k_f^2)-F(k_i^2)]/[k_f^2-k_i^2]$:
\begin{equation}
\begin{array}{l}
X_1(k_f,k_i) = \Delta_{k^2 \sigma_V}(k_f^2,k_i^2)\,,\; \\
X_2(k_f,k_i) = \Delta_{\sigma_V}(k_f^2,k_i^2)\,,\\
%
X_3(k_f,k_i) = \Delta_{\sigma_S}(k_f^2,k_i^2)\,.
\end{array}
\end{equation}

Plainly, in using an \emph{Ansatz} instead of solving the rainbow-ladder Bethe-Salpeter equation for $\chi_\mu(k_f,k_i)$, we expedite progress toward computing the spacelike behaviour of $F_\pi(Q^2)$.  It is a valid procedure so long as nothing essential to understanding the form factor is lost thereby.  This is established by listing the following features of the \emph{Ansatz}.
The first two lines in Eq.\,\eqref{ChiAnsatz} are obtained using the gauge technique \cite{Delbourgo:1977jc}.  Hence, the vertex satisfies the longitudinal Ward-Green-Takahashi (WGT) identity \cite{Ward:1950xp,Green:1953te,Takahashi:1957xn}, is free of kinematic singularities, reduces to the bare vertex in the free-field limit, and has the same Poincar\'e transformation properties as the bare vertex.
With the term in the third line, the \emph{Ansatz} also includes a dressed-quark anomalous magnetic moment, made mandatory by DCSB \cite{Kochelev:1996pv,Bicudo:1998qb,Chang:2010hb,Bashir:2011dp} and the transverse WGT identities \cite{Qin:2013mta}.
Finally, numerical solutions of the rainbow-ladder Bethe-Salpeter equation for the vertex \cite{Maris:1999bh} and algebraic analyses of vertex structure \cite{Chang:2010hb,Bashir:2011dp,Qin:2013mta} show that nonperturbative corrections to the bare vertex are negligible for spacelike momenta $Q^2\gtrsim 1\,$GeV$^2$.
%
A deficiency of Eq.\,\eqref{ChiAnsatz} is omission of nonanalytic structures associated with the $\rho$-meson pole but such features have only a modest impact on $Q^2 r_\pi^2 \lesssim 1$, where $r_\pi$ is the pion's charge radius, and are otherwise immaterial at spacelike momenta \cite{Alkofer:1993gu,Roberts:1994hh,Roberts:2000aa}.

With each of the elements in Eq.\,\eqref{RLFpi} expressed via a generalised spectral representation, as detailed in Ref.\,\cite{Chang:2013pq}, the computation of $F_\pi(Q^2)$ reduces to the act of summing a series of terms, all of which involve a single four-momentum integral.  The integrand denominator in every term is a product of $k$-quadratic forms, each raised to some power.
Within each such term, one employs a Feynman parametrisation in order to combine the denominators into a single quadratic form, raised to the appropriate power.  A suitably chosen change of variables then enables one to readily evaluate the four-momentum integration
using standard algebraic methods.

This is the paramount advantage of our technique: it solves the practical problem of continuing from Euclidean metric to Minkowski space \cite{Bhagwat:2002tx}.  As practitioners continue to find, with gap and Bethe-Salpeter equation solutions represented only by arrays of numbers it is nigh impossible to characterise and track complex-valued singularities that move with increasing $Q^2$ into the domain sampled by a numerical Euclidean-momentum integration, so that choosing and following an acceptable integration contour is practically hopeless.

After calculation of the four-momentum integration, evaluation of the individual term is complete after one computes a finite number of simple integrals; namely, the integrations over Feynman parameters and the spectral integral.
The complete result for $F_\pi(Q^2)$ follows after summing the series.

One aspect of the generalised spectral representations has not yet been explained.  DSE kernels that preserve the one-loop renormalisation group behaviour of QCD will necessarily generate propagators and Bethe-Salpeter amplitudes with a nonzero anomalous dimension $\gamma_F$, where $F$ labels the object concerned.  Consequently, the spectral representation must be capable of describing functions of $\mathpzc{s}=p^2/\Lambda_{\rm QCD}^2$ that exhibit $\ln^{-\gamma_F}[\mathpzc{s}]$ behaviour for $\mathpzc{s}\gg 1$.  This is readily achieved by noting that
\begin{equation}
\ln^{-\gamma_F} [D(\mathpzc{s})]
= \frac{1}{\Gamma(\gamma_F)} \int_0^\infty \! dx\, x^{\gamma_F-1}
\frac{1}{[D(\mathpzc{s})]^x}\,,
\end{equation}
where $D(\mathpzc{s})$ is some function.  Such a factor can be multiplied into any existing spectral representation in order to achieve the required ultraviolet behaviour.  (N.B.\ For practical applications involving convergent four momentum integrals, like those generated by Eq.\,\eqref{RLFpi}, it is adequate to develop and use a power law approximation; viz., $\ln^{\gamma_F} [D(\mathpzc{s})] \approx [D(\mathpzc{s})]^{\mathpzc{p}_F}$.  With ${\mathpzc{p}_F}$ chosen appropriately, this is accurate on the material domain and greatly simplifies the subsequent numerical calculation.)

\begin{figure}[t]
\begin{centering}
\includegraphics[clip,width=0.875\linewidth]{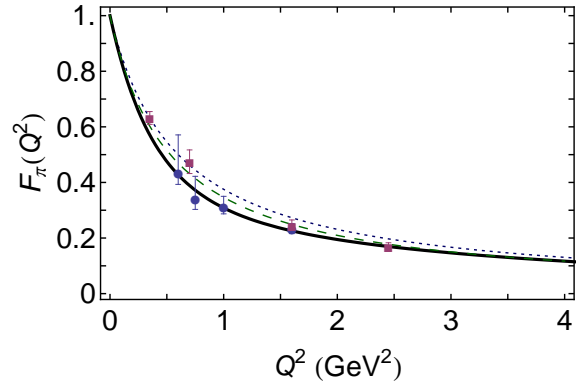}
\end{centering}
\caption{\label{Fig1} \emph{Solid curve} -- Charged pion form factor, computed with $\tilde\eta = 0.5$ in Eq.\,\protect\eqref{ChiAnsatz}; \emph{long-dashed curve} -- calculation in Ref.\,\protect\cite{Maris:2000sk}, which is limited to the domain $Q^2<4\,$GeV$^2$;
and \emph{dotted curve} -- monopole form ``$1/(1+Q^2/m_\rho^2)$,'' where $m_\rho=0.775\,$GeV is the $\rho$-meson mass.  The data are described in Ref.\,\protect\cite{Huber:2008id}.}
\end{figure}

%
\noindent\emph{3:Numerical Results}.\,---\,The pion form factor, computed from Eq.\,\eqref{RLFpi} using the elements and procedures described above, is depicted as curve-A in Fig.\,\ref{Fig1}.
Evidently, this prediction is practically indistinguishable from that described in Ref.\,\cite{Maris:2000sk} on the spacelike domain $Q^2 < 4\,$GeV$^2$, which was the largest value computable reliably in that study.  Critically, however, our prediction extends to arbitrarily large momentum transfers: owing to our improved algorithms, it describes an unambiguous continuation of the earlier DSE prediction to the entire spacelike domain.  It thereby achieves a longstanding goal.

\begin{figure}[t]
\begin{centering}
\includegraphics[clip,width=0.875\linewidth]{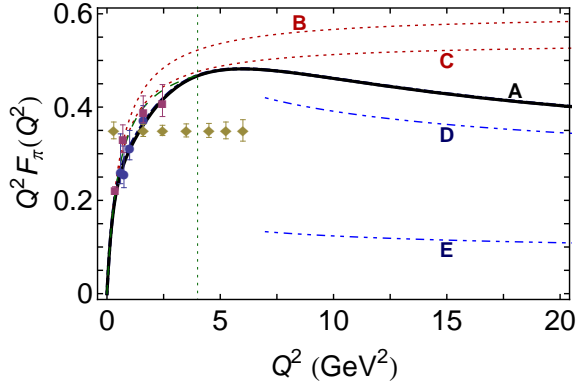}
\end{centering}
\caption{\label{Fig2} $Q^2 F_\pi(Q^2)$.  \emph{Solid curve}\,(A) -- prediction obtained with $\tilde\eta = 0.5$ in Eq.\,\protect\eqref{ChiAnsatz}; \emph{dot-dash curve} -- prediction obtained with $\tilde\eta = 0$; and \emph{long-dashed curve} -- calculation in Ref.\,\protect\cite{Maris:2000sk}, which is limited to the domain $Q^2<4\,$GeV$^2$, whose boundary is indicated by the vertical dotted line.
Remaining curves, from top to bottom: \emph{dotted curve}\,(B) -- monopole form ``$1/(1+Q^2/m_\rho^2)$;'' \emph{dotted curve}\,(C) -- monopole form fitted to data in Ref.\,\protect\cite{Amendolia:1986wj}, with mass-scale $0.74\,$GeV;
\emph{Dot-dot--dashed curve}\,(D) -- Eq.\,\protect\eqref{pionUV} computed with $\varphi_\pi(x)$ in Eq.\,\protect\eqref{phiRLreal}; and \emph{Dot-dot--dashed curve}\,(E) -- Eq.\,\protect\eqref{pionUV} computed with $\varphi_\pi^{\rm asy}(x)$ in Eq.\,\protect\eqref{phiasy}.
The filled-circles and -squares are the data described in Ref.\,\protect\cite{Huber:2008id}; and the filled diamonds indicate the projected reach and accuracy of a forthcoming experiment \protect\cite{E1206101}.
}
\end{figure}

The momentum reach of our improved techniques is emphasised by Fig.\,\ref{Fig2}.   We depict the prediction for $F_\pi(Q^2)$ on the domain $Q^2\in [0,20]\,$GeV$^2$ but have computed the result to $Q^2=100\,$GeV$^2$.  If it were necessary, reliable results could readily be obtained at even higher values.  That is not required, however, because the longstanding questions revolving around $F_\pi(Q^2)$, which we described at the outset, may be answered via Fig.\,\ref{Fig2}.

Before tackling those issues it is important to note that using $\tilde\eta=0.5$, a value commensurate with contemporary estimates \cite{Chang:2010hb,Bashir:2011dp,Qin:2013mta,Chang:2011tx}, the dressed-quark anomalous magnetic moment term in Eq.\,\eqref{ChiAnsatz} has almost no impact on $F_\pi(Q^2)$: the solid and dot-dashed curves in Fig.\,\ref{Fig2} are essentially indistinguishable.  Indeed, for $Q^2 > 4\,$GeV$^2$ there is no difference and hence, as promised in connection with Eq.\,\eqref{ChiAnsatz}, the dressed-quark anomalous magnetic moment has no bearing on the ultraviolet behaviour of the form factor.  On the other hand, it does modestly influence the pion's charge radius: $r_\pi = 0.64\,$fm with $\tilde\eta=0$; whereas $r_\pi = 0.66\,$fm with $\tilde\eta=0.5$.  (Empirically \cite{Beringer:1900zz}, $r_\pi = 0.672 \pm 0.008\,$fm.)  Notably, the radius continues to grow with increasing $\tilde\eta$.  Thus, even though the pion is a pseudoscalar, the dressed-quark anomalous magnetic moment alters the pion's charge distribution.  This effect may be understood as the result of spin-orbit repulsion between the dressed-quarks within the pion, whose rest-frame wave-function necessarily has $P$-wave components in a Poincar\'e-covariant framework \cite{Bhagwat:2006xi}.

We have stressed that the ultraviolet behaviour of $F_\pi(Q^2)$ is of great contemporary interest.  A key feature of the rainbow-ladder prediction for $Q^2 F_\pi(Q^2)$ in Fig.\,\ref{Fig2} is therefore the maximum at $Q^2\approx 6\,$GeV$^2$.  The domain upon which the flattening of the curve associated with this extremum is predicted to occur will be accessible to next-generation experiments \cite{E1206101}.  Unfortunately, on this domain it will still be difficult to distinguish between our prediction and the monopole fitted to data in Ref.\,\cite{Amendolia:1986wj}.

\noindent\emph{4:Drawing connections with perturbative QCD}.\,---\,A maximum appears necessary if $Q^2 F_\pi(Q^2)$ is ever to approach the value predicted by pQCD, Eq.\,\eqref{pionUV}.  In this connection, too, our study has something to add.  The result in Eq.\,\eqref{pionUV4} is associated with curve-E in Fig.\,\ref{Fig2}, which is typically plotted in such figures and described as the prediction of pQCD.  That would be true if, and only if, the pion's valence-quark distribution amplitude were well described by $\varphi_\pi^{\rm asy}(x)$ at the scale $Q^2\sim 4\,$GeV$^2$.  However, that is not the case \cite{Cloet:2013tta}.

The correct comparison with pQCD should be drawn as follows.  Using precisely the interaction that we've employed herein to compute $F_\pi(Q^2)$, one obtains the rainbow-ladder truncation result \cite{Chang:2013pq,Cloet:2013tta}
\begin{equation}
\label{phiRLreal}
\varphi_\pi(x;Q^2=4\,{\rm GeV}^2)
\approx N_{\mathpzc{p}}\, x^{\mathpzc{p}} (1-x)^{\mathpzc{p}}\,,
\end{equation}
with $\mathpzc{p} = 0.3$ and $N_\mathpzc{p} = \Gamma(2(\mathpzc{p}+1))/[\Gamma(\mathpzc{p}+1)]^2$.  This is the amplitude which should be used to calculate the pQCD prediction appropriate for comparison with contemporary experiments.  We depict that computed result as curve-D in Fig.\,\ref{Fig2};\footnote{One might also include the $Q^2$-evolution of $\varphi_\pi(x;Q^2)$ in curve-D.  However, nonperturbative evolution is slow, being overestimated using the leading-order formula, so that ``freezing'' $\varphi_\pi(x;Q^2)=\varphi_\pi(x;Q^2=4\,{\rm GeV}^2)$ provides a valid approximation on the domain depicted.  However, the $Q^2$-evolution of $\varphi_\pi(x;Q^2)$ must be included in a figure that extends to significantly larger $Q^2$ so that the computed approach of curve-D to curve-E is manifest.
}
i.e., this curve is the pQCD prediction obtained when Eq.\,\eqref{phiRLreal} is used in Eqs.\,\eqref{pionUV}--\eqref{wphi}.

Stated simply, curve-D in Fig.\,\ref{Fig2} is the pQCD prediction obtained when the pion valence-quark PDA has the form appropriate to the scale accessible in modern experiments.  Its magnitude is markedly different from that obtained using the asymptotic PDA in Eq.\,\eqref{phiasy}; viz., curve-E, which is only valid at truly asymptotic momenta.
%
The meaning of ``truly asymptotic'' is readily illustrated.  The PDA in Eq.\,\eqref{phiRLreal} produces $\mathpzc{w}_\varphi^2 = 3.2$, which is to be compared with the value computed using the asymptotic PDA: $\mathpzc{w}^{\rm asy}_\varphi = 1.0$.  Applying leading-order QCD evolution to the PDA in Eq.\,\eqref{phiRLreal}, one must reach momentum transfer scales $Q^2 > 1000\,$GeV$^2$ before $\mathpzc{w}_\varphi^2 < 1.6$; i.e., before $\mathpzc{w}_\varphi^2$ falls below half its original value.

%
\noindent\emph{5:Summary}.\,---\,Given the observations above, the near agreement between the pertinent perturbative QCD prediction in Fig.\,\ref{Fig2} (curve-D) and our predicted form of $Q^2 F_\pi(Q^2)$ (curve-A) is striking.  It highlights that a single DSE interaction kernel, determined fully by just one parameter and preserving the one-loop renormalisation group behaviour of QCD, has completed the task of unifying the pion's electromagnetic form factor and its valence-quark distribution amplitude; and, indeed, numerous other quantities \cite{Chang:2011vu,Bashir:2012fs,Maris:2003vk,Eichmann:2011ej}.

Moreover, this leading-order, leading-twist QCD prediction, obtained with a pion valence-quark PDA evaluated at a scale appropriate to the experiment, Eq.\,\eqref{phiRLreal}, underestimates our full computation by merely an approximately uniform 15\% on the domain depicted.
%
%
%
The small mismatch is not eliminated by variation of $\Lambda_{\rm QCD}$ within its empirical bounds, which shifts curve-D by only $\pm 3$\% at $Q^2=20\,$GeV$^2$.  It is instead explained by a combination of higher-order, higher-twist corrections to Eq.\,\eqref{pionUV} in pQCD on the one hand, and shortcomings in the rainbow-ladder truncation, which predicts the correct power-law behaviour for the form factor but not precisely the right anomalous dimension in the strong coupling calculation on the other hand.
Hence, as anticipated earlier \cite{Maris:1998hc} (and expressing a result that can be understood via the behaviour of the dressed-quark mass-function \cite{Chang:2011vu,Bashir:2012fs}), one should expect dominance of hard contributions to the pion form factor for $Q^2\gtrsim 8\,$GeV$^2$.  Notwithstanding this, the normalisation of the form factor is fixed by a pion wave-function whose dilation with respect to $\varphi_\pi^{\rm asy}(x)$ is a definitive signature of dynamical chiral symmetry breaking, which is such a crucial feature of the Standard Model.


%
Work supported by:
For\-schungs\-zentrum J\"ulich GmbH;
Department of Energy, Office of Nuclear Physics, contract no.~DE-AC02-06CH11357;
and 
National Science Foundation, grant no.\
NSF-PHY1206187.



\end{document}